*Research Article*

# Sex-based Disparities in Brain Aging: A Focus on Parkinson's Disease

Iman Beheshti[1,2], Samuel Booth[1,2], and Ji Hyun Ko[1,2,3,*]

[1]Department of Human Anatomy and Cell Science, University of Manitoba, Winnipeg, MB, Canada

[2]Kleysen Institute for Advanced Medicine, Health Science Centre, Winnipeg, MB, Canada

[3]Graduate Program in Biomedical Engineering, University of Manitoba, Winnipeg, MB, Canada

*Corresponding Author:

Dr. Ji Hyun Ko

Address: 130-745 Bannatyne Ave., Winnipeg, Manitoba, R3E 0J9, Canada

Telephone: +1-204-318-2566

Fax: +1-204-789-3920

Email: ji.ko@umanitoba.ca

# ABSTRACT

## INTRODUCTION:

Parkinson's disease (PD) is linked to faster brain aging. Sex is recognized as an important factor in PD, such that males are twice as likely as females to have the disease and have more severe symptoms and a faster progression rate. Despite previous research, there remains a significant gap in understanding the function of sex in the process of brain aging in PD patients.

## METHODS:

The T1-weighted MRI-driven brain-predicted age difference (i.e., brain-PAD: the actual age subtracted from the brain-predicted age) was computed in a group of 373 PD patients (mean age ± SD: 61.37 ± 9.81, age range: 33–85, 34% female) from the Parkinson's Progression Marker Initiative database using a robust brain-age estimation framework that was trained on 949 healthy subjects. Linear regression models were used to investigate the association between brain-PAD and clinical variables in PD, stratified by sex. All female PD patients were used in the correlational analysis while the same number of males were selected based on propensity score matching method considering age, education level, age of symptom onset, and clinical symptom severity.

## RESULTS:

Despite both patient groups being matched for demographics, motor and non-motor symptoms, it was observed that males with Parkinson's disease (PD-M) exhibited a significantly higher mean brain age-delta than their female counterparts (PD-F) ($t(256)= 2.50$, $p = 0.012$). In the propensity score-matched PD-M group (PD-M*), brain-PAD was found to be associated with a decline in general cognition, a worse degree of sleep behavior disorder, reduced visuospatial acuity, and caudate atrophy. Conversely, no significant links were observed between these factors and brain-PAD in the PD-F group.

## DISCUSSION:

Having 'older' looking brains in PD-M than PD-F supports the idea that sex plays a vital function in PD, such that the PD mechanism may be different in males and females. This study has the potential to broaden our understanding of dissimilarities in brain aging between sexes in the context of PD.

1. Introduction

Parkinson's disease (PD) is a progressive neurodegenerative disorder that is characterized by both motor and non-motor symptoms [1]. Tremors, rigidity, slowness of movement, and difficulty in walking are the most noticeable motor abnormalities found in PD [2]. In addition, the majority of PD patients also experience cognitive decline (up to 50%), impulse control disorders (up to 60%), apathy/anhedonia (up to 40%), depression (up to 35%), and anxiety (up to 60%) [2]. A growing number of studies have attempted to identify the risk factors in PD. Although the fundamental causes of PD are yet unknown, several factors appear to play a vital role, including genetic predisposition and advanced age. Male sex is also recognized as a significant factor that contributes to the development and phenotypic expression of PD [3]. The prevalence of PD is twice as high in males compared to females, and is frequently associated with earlier disease onset [4]. It has been observed that tremor-dominant PD is more prevalent in females, whereas rigid-dominant PD is more commonly found in males [4]. In the cognitive domain, male patients are more likely to experience mild cognitive impairment and have more rapid progression to dementia [5,6]. Additionally, there are sex differences in the expression of symptoms across different cognitive domains. For example, female PD patients generally perform better than males in verbal cognitive tasks [4]. Males typically experience more significant impairments in verbal fluency, verbal memory, and facial emotion identification, whereas female PD patients typically experience more diminished visuospatial cognition [3].

These findings demonstrate that sex plays a substantial role in the clinical presentation of PD, with women displaying milder symptoms [7]. To date, a few neuroimaging studies have explored sex differences in PD. For instance, a more recent study investigated sex differences in PD using anatomical MRI data, including deformation-based morphometry, cortical thickness, and diffusion-weighted MRI measures, on a large sample of PD patients [7]. Tremblay et al. [7] reported that, when disease duration and severity were equal, male PD patients have more severe brain atrophy in the right postcentral gyrus, bilateral frontal lobes, left insular lobe, left thalamus, left inferior temporal gyrus, and cingulate gyrus. On the contrary, female patients have more severe tissue loss in the right occipital cortex, the left frontal lobe, the left insular gyrus, and the right parietal lobe [7]. Similarly, reduced cortical thickness has been observed in the frontal, temporal, occipital, and parietal lobes in male PD patients than their female counterparts [8].

Old age is one of the most significant risk factors for PD. It has been postulated that the normal ageing process is exaggerated in specific vulnerable brain regions in PD [9], including the substantia nigra, while earlier neuroimaging studies reported a divergent whole-brain metabolic pattern in normal aging and PD [10]. Recent advances in machine learning (ML) techniques have made it possible to estimate an individual's "Brain Age" more accurately using their neuroimaging data [11]. The typical output of a brain age estimation model is known as Brain-PAD (i.e., brain-predicted age minus actual age), which can quantify the degree of global brain health [11]. The Brain-PAD biomarker has found widespread application in the identification of accelerated brain aging in a range of neurological disorders, such as Alzheimer's disease [12], epilepsy [13], and schizophrenia [14].

To date, only a few studies have investigated the brain age estimation technique in the area of PD [15,16]. These studies mostly focused on quantifying Brain-PAD and the association between Brain-PAD scores and clinical variables in PD. For instance, a strong correlation between brain-PAD scores and Unified Parkinson's Disease Rating Scale (UPDRS) III scores (motor symptom severity) as well as a weak correlation between brain-PAD scores and Montreal Cognitive Assessment (MoCA) scores has been reported in PD [15,16]. However, investigation into sex differences in PD using brain age measures has not been performed. In this study, we investigated whether clinical symptom severity correlates differently with Brain-PAD in male versus female PD patients.

## 2. Material and methods

### 2.1 Participants and MRI Acquisition

To train and validate a model that estimate Brain Age, we employed a total of 1,054 T1-weighted (T1w) magnetic resonance imaging (MRI) scans from healthy controls (HCs) obtained from the Open Access Series of Imaging Studies (OASIS) (https://www.oasis-brains.org/), the IXI (http://brain-development.org/ixi-dataset/), and the Parkinson's Progression Markers Initiative (PPMI) (www.ppmi-info.org) databases. All healthy controls were free from cognitive impairment or any neurological diseases as per the databases. A brain age estimation model was trained using a randomly selected 90% of healthy controls (HCs), consisting of 949 individuals with a mean age of 49.75±18.96 years and an age range of 18-94 years, of which 54% were female. The remaining 10% of HCs, totaling 105 individuals with a mean age of 48.62±19.14 years and an age range of 18-93 years, of which 53% were female, were held out as a sample for validation purposes.

  To test whether Brain Age was differently correlated with clinical symptoms severities between sexes, a total of 373 individuals diagnosed with PD were included in the study, with a mean age of 61.37±9.81 years and an age range of 33 to 85 years. Among the PD participants, 34% were female. Baseline T1w MRI scans and clinical data were obtained from the PPMI database in September 2022, and brain age was calculated for each subject.

  In addition to demographic characteristics (i.e., age, education, onset age, and disease duration), we collected clinical measures including motor symptom scores (i.e., UPDRS-III (total),

UPDRS-III (total rigidity), and UPDRS-III (total tremor)), non-motor symptom scores (i.e., MoCA, Epworth Sleepiness Scale (ESS), Letter Number Sequencing, Sleep Behavior Disorder Question Score (REM), Benton Judgment of Line Orientation Score (BJLO), Hopkins Verbal Learning Test (HVLT) delayed recall, HVLT delayed recognition, olfactory testing, Symbol Digit Modalities Score), mood (i.e., anxiety and Geriatric Depression Scale (GDS)), cerebrospinal fluid (CSF) biomarkers (i.e., alpha-synuclein (a-syn), amyloid-beta, and CSF p-tau (2016 assay)) for PD patients. We also included measurements of the left and right Striatal Binding Ratios (SBRs) derived from DaTScan Single-Photon Emission Computed Tomography (SPECT) for the caudate and putamen regions. SBR values were obtained from the PPMI database. Briefly, SPECT data underwent HOSEM reconstruction in the HERMES system without filtering. Reconstructed files were processed in PMOD. Images underwent attenuation correction and a Gaussian 3D filter. They were normalized to the Montreal Neurologic Institute (MNI) space. Regions of interest (ROIs) were defined for the caudate, putamen, and occipital cortex (reference tissue). SBR was calculated by subtracting one from the ratio of the target to reference regions. Detailed information regarding SBR contribution can be found at https://www.ppmi-info.org.

In the group under study, the number of individuals diagnosed with PD consisted of 244 males and 129 females. To ensure that the severity of disease was consistent across both sexes in the PD group, we identified a set of male PD patients (PD-M*, N = 129) through propensity score matching from the larger group of 244 male patients. These patients were matched based on actual age, education level, age of symptom onset and diagnosis, UPDRS-III (total), UPDRS-III (rigidity), UPDRS-III (tremor), MoCA, REM, and ESS scores. Propensity score matching was conducted using the *pymatch* package in Python (https://github.com/benmiroglio/pymatch), which employs logistic regression models to generate propensity scores and match two groups.

**2-2 Image Processing and Brain Age Estimation Model**

Brain age estimation was performed based on T1-weighted MRI data. The voxel-based morphometry (VBM) technique implemented in CAT12 toolbox (http://www.neuro.uni-jena.de/cat/), as an extension of the Statistical Parametric Mapping (SPM12) software package (https://www.fil.ion.ucl.ac.uk/spm/software/spm12/), was used for preprocessing T1-weighted MRI scans with the default set of parameters, including DARTEL normalization and modulation for nonlinear components, as described in [17]. Based on the VBM technique, we generated the density images of gray matter (GM), white matter (WM), and cerebrospinal fluid (CSF) for each T1w MRI scan. We then smoothed these images with a 4-mm kernel. The smoothed GM, WM, and CSF images were re-sampled to an 8-mm isotropic spatial resolution, resulting in 3,747 voxels per volume. We also computed the total volumes GM, WM, and CSF for each subject using the CAT12 toolbox as well as the total intracranial volume (TIV). This procedure was applied to both the training and testing datasets. Visual assessment of the quality of MRI processing and segmentation was done by IB, with a second quality assurance performed via the "Check homogeneity function" in CAT12. To estimate the values of brain age, we utilized a support vector regression (SVR) algorithm with a linear kernel that was implemented in MATLAB_R2020b (The Mathworks, Natick, MA, USA).

In the prediction model, the independent variables comprised of the voxel intensities of GM, WM, and CSF, along with the variables of sex, scanner vendor, field strength, TIV, and total brain volumes of GM, WM, and CSF (Table 1). The dependent variable was the actual age. The mean absolute error (MAE), root mean square error (RMSE), and coefficient of determination ($R^2$) between actual age and model-estimated age were used to assess prediction performance in the training set (N = 949) using a 10-fold cross-validation strategy, as well as in the hold-out control set (N = 105). The brain-PAD (i.e., model-estimated age minus actual age) was also calculated in the form of the mean and a 95% confidence interval (CI). The bias-free brain age values were computed using a validated bias adjustment scheme, which is detailed in [18]. The bias adjustment approach utilized in this study is publicly available at: https://github.com/medicslab/Bias_Correction. Next, the final prediction model was developed using the entire training set (N=949) and applied to hold-out sets (HC: N=105; PD: N=373) to compute the brain-PAD for those sets. A brain-PAD value close to zero (i.e., estimated age ≅ actual age) stands for the point that the subject being studied follows a healthy brain aging trajectory. A negative brain-PAD value (i.e., estimated age < actual age) indicates a younger-looking brain, and a positive brain-PAD value (i.e., estimated age > actual age) indicates an older-looking brain.

Table 1: List of Independent Variables used in a Brain Age Prediction Model.

| Variable | Description | Number of attributes |
|---|---|---|
| Voxel intensities of GM | Intensity values of gray matter voxels | 3,747 |
| Voxel intensities of WM | Intensity values of white matter voxels | 3,747 |
| Voxel intensities of CSF | Intensity values of cerebrospinal fluid voxels | 3,747 |
| Total brain volume of GM | Total volume of gray matter in the brain | 1 |
| Total brain volume of WM | Total volume of white matter in the brain | 1 |
| Total brain volume of CSF | Total volume of cerebrospinal fluid in the brain | 1 |
| Total intracranial volume | Total volume of the brain | 1 |
| Sex | Biological sex of the participant | 1 |
| Scanner vendor | Manufacturer or brand of the scanning equipment | 1 |
| Field strength | Magnetic field strength of the scanner | 1 |

## 2.3 Statistical analysis

The mean brain-PAD between the hold-out sets was examined using an independent Student's t-test. We used multiple linear regression models to examine whether brain-PAD is able to predict the clinical variables in PD. The models that have been tested are:

1) Motor symptom severity (UPDRS-III) ~ 1 + brain-PAD + age + disease duration + education + sex

2) Non-motor symptom severity (see Table 2) ~ 1 + brain-PAD + age + disease duration + education + sex + UPDRS-III

UPDRS-III was considered in equation (2) to examine whether it influences the severities of non-motor symptoms. For mood symptoms, the severity of cognitive symptoms assessed by the MoCA was also modeled:

3) Mood symptom severity (anxiety (i.e., the state-trait anxiety inventory test) and GDS) ~ 1 + brain-PAD + age + disease duration + education + sex + UPDRS-III + MoCA

We repeated the above regression models for males and females separately to assess the differential association of brain-PAD and clinical symptom severity in each sex. Likewise, UPDRS-III and MoCA were considered in equation (3) to examine whether motor and non-motor symptoms influence mood status in PD. All statistical analyses were performed in MATLAB using the *regstats* function, and a P-value of less than 0.05 was considered significant. For each statistical analysis, we excluded subjects with missing data.

To visualize the brain regions that were specifically associated advanced brain age, we used multiple regression analysis in SPM12 to predict GM/WM volumes with Brain-PAD scores as an independent variable in each group (i.e., HC and PD), separately. To identify the brain regions that were differently correlated with Brain-PAD scores in PD vs. HC, regression coefficients were contrasted between the groups. For these analyses, the peak-level p-values were considered significant at .05 after corrected with a family-wise error (FWE). We only considered clusters with > 100 voxels. All regression analyses incorporated age, sex, and TIV as covariates.

Table 2: Demographics and clinical characteristics of PD patients used in this study by sex.

| Characteristics | # | PD (N = 373) | | PD-M (N = 244) | | PD-F (N = 129) | | PD-M* (N = 129) | | $P^*$ | $P^{**}$ |
|---|---|---|---|---|---|---|---|---|---|---|---|
| **Demographics** | | mean | SD | mean | SD | mean | SD | mean | SD | | |
| Age, years | 0 | 61.37 | 9.81 | 62.04 | 9.87 | 60.11 | 9.63 | 61.14 | 9.93 | 0.07 | 0.40 |
| Onset age | 0 | 59.34 | 10.06 | 60.06 | 10.11 | 57.99 | 9.84 | 59.17 | 10.27 | 0.06 | 0.35 |
| Education, years | 0 | 15.55 | 2.88 | 15.74 | 2.88 | 15.19 | 2.84 | 15.50 | 2.75 | 0.08 | 0.37 |
| Disease duration, months | 0 | 6.73 | 6.58 | 6.42 | 6.05 | 7.32 | 7.48 | 6.32 | 5.73 | 0.21 | 0.23 |
| **Motor symptoms** | | | | | | | | | | | |
| UPDRS-III (total) | 1 | 32.22 | 13.21 | 32.52 | 13.34 | 31.66 | 12.99 | 30.13 | 13.22 | 0.55 | 0.35 |
| UPDRS-III (total rigidity) | 0 | 3.83 | 2.63 | 4.13 | 2.67 | 3.26 | 2.46 | 3.63 | 2.62 | 0.00 | 0.25 |
| UPDRS-III (total tremor) | 0 | 4.34 | 3.16 | 4.37 | 3.18 | 4.27 | 3.13 | 4.28 | 3.17 | 0.77 | 0.98 |
| **Non-motor symptoms** | | | | | | | | | | | |
| MoCA | 0 | 27.07 | 2.32 | 26.88 | 2.34 | 27.43 | 2.25 | 27.10 | 2.32 | 0.03 | 0.24 |
| Epworth sleepiness scale | 0 | 5.65 | 3.39 | 5.73 | 3.27 | 5.50 | 3.61 | 5.60 | 3.31 | 0.53 | 0.80 |
| Letter Number Sequencing Score | 0 | 10.56 | 2.66 | 10.45 | 2.72 | 10.78 | 2.54 | 10.48 | 2.92 | 0.26 | 0.39 |
| REM | 2 | 4.08 | 2.64 | 4.20 | 2.71 | 3.85 | 2.50 | 3.85 | 2.69 | 0.22 | 1.00 |
| BJLO | 0 | 12.81 | 2.11 | 13.15 | 2.02 | 12.17 | 2.13 | 12.70 | 2.17 | 0.00 | 0.06 |
| HVLT Delayed Recall | 0 | 8.36 | 2.50 | 8.00 | 2.55 | 9.06 | 2.25 | 8.15 | 2.49 | 0.00 | 0.00 |
| HVLT Delayed Recognition | 1 | 11.22 | 1.09 | 11.12 | 1.14 | 11.40 | 0.96 | 11.21 | 1.04 | 0.02 | 0.12 |
| Olfactory testing | 0 | 22.49 | 8.33 | 21.44 | 8.18 | 24.48 | 8.27 | 21.80 | 8.82 | 0.00 | 0.01 |
| Symbol Digit Modalities Score | 0 | 41.48 | 9.83 | 40.19 | 9.77 | 43.93 | 9.51 | 40.78 | 9.21 | 0.00 | 0.01 |
| **Mood** | | | | | | | | | | | |
| Anxiety | 1 | 65.27 | 18.06 | 64.21 | 17.59 | 67.30 | 18.81 | 64.84 | 17.99 | 0.12 | 0.28 |
| GDS | 0 | 2.29 | 2.38 | 2.32 | 2.35 | 2.24 | 2.43 | 2.30 | 2.40 | 0.76 | 0.84 |
| **CSF Biomarkers, pg/ml** | | | | | | | | | | | |
| a-Synuclein (a-syn) | 9 | 1502 | 669 | 1458 | 647 | 1585 | 702 | 1448 | 607 | 0.08 | 0.10 |

| | | | | | | | | | | |
|---|---|---|---|---|---|---|---|---|---|---|
| Amyloid-b | 13 | 918 | 416 | 901 | 377 | 950 | 481 | 904 | 364 | 0.29 | 0.39 |
| CSF p-tau (2016 assay) | 39 | 14.78 | 5.19 | 14.61 | 4.73 | 15.10 | 5.97 | 14.43 | 4.88 | 0.42 | 0.35 |
| **Other variables** | | | | | | | | | | | |
| CAUDATE (L+R) | 3 | 4.01 | 1.11 | 3.95 | 1.08 | 4.12 | 1.16 | 4.04 | 1.09 | 0.17 | 0.59 |
| PUTAMEN (L+R) | 3 | 1.64 | 0.58 | 1.62 | 0.55 | 1.68 | 0.63 | 1.72 | 0.55 | 0.40 | 0.58 |

PD: Parkinson's disease; F: females; M: males; M*: matched males; UPDRS: Unified Parkinson's Disease Rating Scale; MoCA: Montreal Cognitive Assessment; REM: Sleep Behavior Disorder Questionnaire Score; BJLO: Benton Judgement of Line Orientation Score; HVLT: Hopkins Verbal Learning Test; GDS: Geriatric Depression Scale; N: number of subjects; #: number of missing values in each variable; L: left; R: right; P*: t-test between PD-F and PD-M; P**: t-test between PD-F and PD-M*.

## 3. Results

### 3.1 Sex differences in clinical and biomarker scores

All diagnostic assessments and symptom ratings were obtained from the PPMI dataset. The demographic characteristics, including age, education, onset age, and disease duration, were similar between female and male PD patients (Table 2). While the overall severity of motor symptoms (total UPDRS-III) was not significantly different between males and females ($t(370)=0.59$, $p=0.551$), the rigidity subscale was significantly higher in male PD patients ($t(371) = 3.04$, $p = 0.002$). The tremor subscale was not significantly different between sexes ($t(371)= 0.29$, $p=0.767$).

Significant differences were observed among cognitive performance scores as measured by MoCA ($t(371) = 2.20$, $p = 0.028$), BJLO ($t(371) = 4.38$, $p < 0.001$), HVLT Delayed Recall ($t(371) = 4.00$, $p < 0.001$), HVLT Delayed Recognition ($t(370) = 2.40$, $p = 0.016$), olfactory testing ($t(371) = 3.40$, $p < 0.001$), and Symbol Digit Modalities Score ($t(371) = 3.54$, $p < 0.001$). There was no significant difference between sexes in the Epworth Sleepiness Scale ($t(371) = 0.63$, $p = 0.527$), Letter Number Sequencing Score ($t(371) = 1.11$, $p = 0.264$), and REM ($t(369) = 1.21$, $p = 0.224$). Anxiety ($t(370) = 1.57$, $p = 0.116$) and depression scores (GDS: $t(371)= 0.30$, $p =0.759$) were not significantly different between sexes. No significant differences were observed in CSF biomarkers (i.e., α-Synuclein ($t(362) = 1.73$, $p = 0.083$), Amyloid-β ($t(358)= 1.05$, $p =0.291$), and CSF p-tau ($t(332) = 0.80$, $p = 0.418$)). Regional atrophy in caudate ($t(368) = 1.39$, $p = 0.165$) and putamen ($t(368)= 0.84$, $p = 0.398$) were not significantly different, either.

To control for different clinical symptom severity and other demographic variables in the correlational analyses below, we used propensity score matching method to select 129 male PD patients (PD-M*). We confirmed that there was no significant difference between PD-M* and PD-F in age (t(256)= 0.84, p = 0.398), education level (t(256)= 0.89, p = 0.373), age of symptom onset (t(256)= 0.949, p = 0.345), age of diagnosis (t(256)= 0.921, p = 0.357), UPDRS-III (total) (t(256)= 0.932, p = 0.352), UPDRS-III (total rigidity) (t(256)= 1.156, p = 0.251), UPDRS-III (total tremor) (t(256)= 0.019, p = 0.984), MoCA (t(256)= 1.17, p = 0.242), REM (t(256)= 0.00, p = 1), and ESS (t(256)= 0.251, p = 0.801) (Table 2).

**3-2 Brain age across different groups**

The MAE in the training set (achieved through a 10-fold cross-validation strategy) and the hold-out HC set were 4.72 years and 4.63 years, respectively, which were well comparable with those of other studies [13,19]. The mean of brain-PAD in the training and hold-out HC sets was approximately zero (Table 3). There was no significant difference between males and females in the training set (t(947) = 0.52, p = 0.60) as well as the hold-out HC (t(103) = 0.56, p = 0.57) in brain-PAD. As expected, the PD patients showed a positive mean brain-PAD of 2.92 [95% CI 2.20-3.65] years, which was significantly higher than that of the hold-out HC (t(476) = 3.94, p< 0.0001). Our mean brain-PAD of 2.92 years is similar to what has been reported in PD [15,16]. Fig 1 displays the association between estimated ages versus the actual ages in different groups. Regression parameters were almost identical between training and hold-out sets of HCs, while PD patients showed higher intersection (+4.3 years) than the identity line (t(140) = 5.00, p< 0.001). Fig 2 shows the contrasts of brain-PAD in different subsets. All PD subgroups showed higher brain-PAD than HC (hold-out set) (p<0.05). PD-M patients had a significantly higher brain-PAD than PD-F patients (PD-M: 3.53, 95% CI 2.61-4.45 years; PD-F: 1.78, 95% CI 0.59-2.96 years; t(371) = 2.26, p = 0.024) (Table 3). The significance was preserved when PD-M* was compared with PD-F patients (PD-M*: 3.85, 95% CI 2.72-4.95 years; PD-F: 1.78, 95% CI 0.59-2.96 years; t(256) = 2.50, P = 0.012).

Table 3: The Results of Brain Age Values on Different Sets.

| Set | Group | N | MAE (yrs) | RMSE (yrs) | R2 | Mean Brain-PAD (yrs) | 95% CI Values (yrs) |
|---|---|---|---|---|---|---|---|
| Training set | HC | 949 | 4.72 | 6.07 | 0.91 | 0.00 | -0.38, 0.38 |
| Hold-out set | HC | 105 | 4.63 | 5.88 | 0.91 | -0.08 | -1.23, 1.05 |
| Test set | PD (all) | 373 | 5.96 | 7.72 | 0.64 | 2.92 | 2.20, 3.65 |
| Test set | PD-F | 129 | 5.30 | 7.00 | 0.63 | 1.78 | 0.59, 2.96 |
| Test set | PD-M | 244 | 6.31 | 8.08 | 0.64 | 3.53 | 2.61, 4.45 |
| Test set | PD-M* | 129 | 6.02 | 7.49 | 0.73 | 3.85 | 2.72, 4.95 |

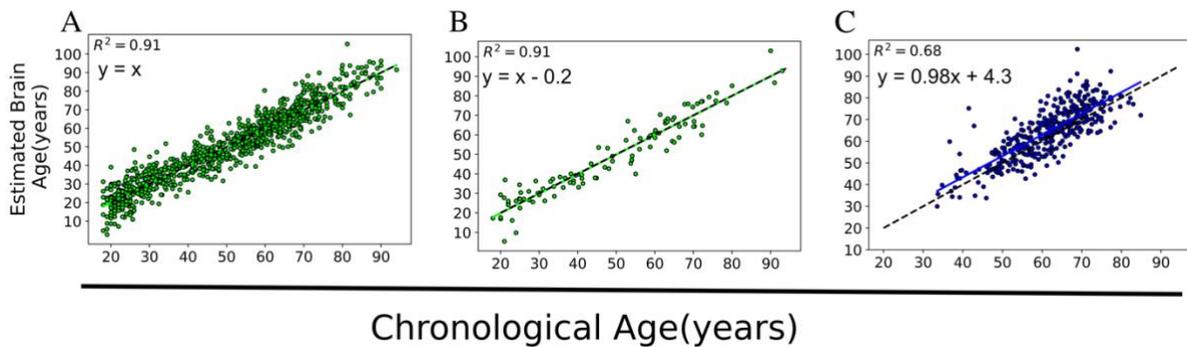

**Fig 1:** Scatter plots showing the estimated brain age versus actual age in different datasets. A) The training set (N = 949) that was evaluated through a 10-fold cross-validation strategy, B) the hold-out HC (N = 105), and C) PD (N = 373). The dashed black line represents the identity line (y = x).

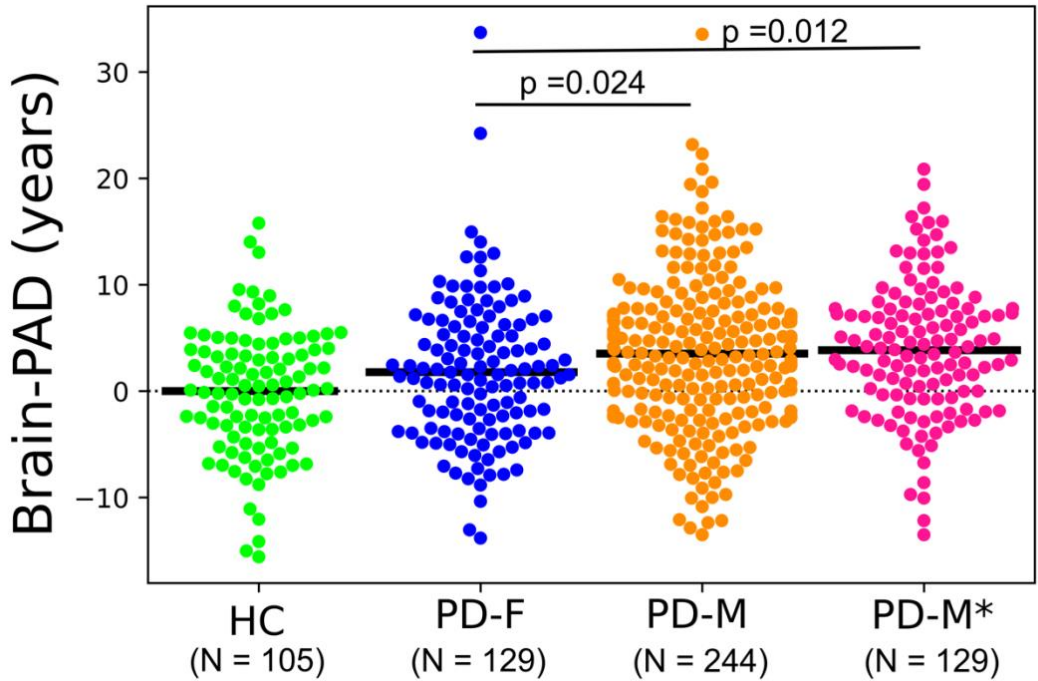

**Fig 2.** Box-plots showing the grouped Brain-PAD values among the hold-out HC and PD patients with respect to the sex categories. The solid black line represents the mean brain-PAD values of each group, while the dashed black line represents the reference line (y = 0). The statistical tests between groups were performed using a student t-test.

### 3.3. Brain age and motor symptoms

Multiple linear regression analysis showed that increased UPDRS-III scores in the PD group were associated with higher brain-PAD (t(366) = 2.81, p = 0.005), actual age (t(366) = 3.33, p = 0.001), and duration of the disease (t(366) = 2.56, p = 0.011) (Table 4). This pattern was similar in the PD-M group, but an increase in UPDRS-III (total) scores in the PD-M* group was associated only with a higher actual age (t(124) = 2.61, p = 0.010). UPDRS-III in the PD-F group were not associated with brain-PAD (t(123) = 0.939, p = 0.35), actual age (t(123) = -0.238, p = 0.81), disease duration (t(123) = 1.48, p = 0.14), and education (t(123) = -1.44, p = 0.15). No association was observed between brain-PAD and the UPDRS-III (total rigidity) or between brain-PAD and the UPDRS-III (total tremor) in our investigation.

Table 4: Regression Model Outputs in PD: Motor Symptoms.

| Model | | UPDRS-III (total) | | | | UPDRS-III (total rigidity) | | | | UPDRS-III (total tremor) | | | |
|---|---|---|---|---|---|---|---|---|---|---|---|---|---|
| | | PD | PD-F | PD-M | PD-M* | PD | PD-F | PD-M | PD-M* | PD | PD-F | PD-M | PD-M* |
| Brain-PAD | b | 0.265 | 0.159 | 0.295 | 0.182 | 0.031 | 0.020 | 0.033 | 0.007 | -0.016 | -0.056 | 0.001 | 0.022 |
| | t | 2.815 | 0.939 | 2.641 | 1.047 | 1.640 | 0.637 | 1.414 | 0.181 | -0.712 | -1.380 | 0.031 | 0.510 |
| | p | **0.005** | 0.350 | **0.009** | 0.297 | 0.102 | 0.525 | 0.159 | 0.857 | 0.477 | 0.170 | 0.976 | 0.611 |
| Actual age | b | 0.230 | -0.029 | 0.341 | 0.302 | 0.002 | -0.024 | 0.014 | 0.011 | 0.058 | 0.038 | 0.068 | 0.069 |
| | t | 3.337 | -0.238 | 4.112 | 2.617 | 0.151 | -1.052 | 0.808 | 0.454 | 3.545 | 1.337 | 3.349 | 2.450 |
| | p | **0.001** | 0.812 | **0.000** | **0.010** | 0.880 | 0.295 | 0.420 | 0.651 | **0.000** | 0.184 | **0.001** | **0.016** |
| Duration | b | 0.264 | 0.232 | 0.283 | 0.295 | 0.025 | 0.057 | -0.001 | 0.023 | 0.057 | 0.057 | 0.060 | 0.097 |
| | t | 2.565 | 1.482 | 2.090 | 1.464 | 1.189 | 1.967 | -0.045 | 0.536 | 2.321 | 1.542 | 1.796 | 1.990 |
| | p | **0.011** | 0.141 | **0.038** | 0.146 | 0.235 | 0.051 | 0.964 | 0.593 | **0.021** | 0.126 | 0.074 | **0.049** |
| Education | b | 0.007 | -0.584 | 0.250 | 0.540 | 0.019 | -0.010 | 0.032 | 0.084 | 0.082 | 0.135 | 0.050 | -0.007 |
| | t | 0.030 | -1.446 | 0.884 | 1.310 | 0.408 | -0.130 | 0.542 | 0.975 | 1.469 | 1.406 | 0.727 | -0.067 |
| | p | 0.976 | 0.151 | 0.377 | 0.193 | 0.683 | 0.897 | 0.588 | 0.331 | 0.143 | 0.162 | 0.468 | 0.947 |
| Sex | b | -0.121 | - | - | - | -0.816 | - | - | - | -0.023 | - | - | - |
| | t | -0.085 | - | - | - | -2.833 | - | - | - | -0.068 | - | - | - |
| | p | 0.932 | - | - | - | **0.005** | - | - | - | 0.946 | - | - | - |

PD, Parkinson diseases; F, females; M, males; M*, matched males; UPDRS, Unified Parkinson Disease Rating Scale. Significant results are highlighted in bold.

### 3.4. Brain age and cognitive symptoms

As expected, lower MoCA scores in the PD group were associated with a higher Actual age (t(365) = -3.78, p < 0.001) and lower education (t(365) = 2.05, p = 0.041) (Table 5). While only non-significant trend-level correlation was observed between Brain-PAD when the whole PD patients were considered (t(365)=-1.818, p=0.070), lower MoCA scores were associated with higher brain-PAD (t(238) = -2.01, p = 0.045) when male PD patients were separately analyzed (PD-M). The association between MoCA and brain-PAD remained to be significant in the PD-M* group (t(123)= -2.14, p = 0.034), while PD-F group showed no significant associations between MoCA and brain-PAD (t(122) = -2.27, p = 0.78).

Epworth Sleepiness Scale scores were not associated with Brain-PAD. The Letter Number Sequencing Score was observed to be associated with brain-PAD only when all male and female patients were combined (t(365)=-2.074, p=0.039). A positive association between REM and brain-PAD was observed in PD (t(363) = 3.09, P = 0.002), PD-M (t(237)= 3.15, p = 0.002), and PD-M* (t(122) = 3.53, p = 0.001) groups. However, REM were not observed to be associated with brain-PAD in the PD-F group (t(121)= 0.88, p = 0.37). REM was strongly positively associated with UPDRS-III (total) in all groups (Table 5). Decreased BJLO scores were associated with higher brain-PAD in the PD (t(365) = -2.36, p = 0.019) and PD-M* (t(123) = -2.17, p = 0.032) groups, but not in the PD-F (t(122) = -1.57, p = 0.119) and PD-M ((238)t = -1.76, p = 0.079) groups. Higher BJLO scores were found to be associated with higher education in all groups. Brain-PAD was not significantly associated with HVLT Delayed Recall scores, HVLT Delayed Recognition, olfactory testing, and Symbol Digit Modalities Score. Instead, actual age was correlated with HVLT Delayed Recall scores, HVLT Delayed Recognition, olfactory testing, and Symbol Digit Modalities Score in all subgroups (p<0.02). Detailed of these multiple linear regression analyses are depicted in Table 5.

Table 5: Regression Model Outputs in PD: Cognitive Symptoms.

| Model | | MoCA | | | | Epworth sleepiness scale score | | | | Letter Number Sequencing Score | | | |
|---|---|---|---|---|---|---|---|---|---|---|---|---|---|
| | | PD | PD-F | PD-M | PD-M* | PD | PD-F | PD-M | PD-M* | PD | PD-F | PD-M | PD-M* |
| Brain-PAD | b | -0.030 | -0.008 | -0.040 | -0.065 | 0.007 | -0.036 | 0.028 | 0.062 | -0.037 | -0.042 | -0.033 | -0.065 |
| | t | -1.818 | -0.278 | -2.016 | -2.143 | 0.279 | -0.767 | 1.001 | 1.385 | -2.074 | -1.310 | -1.508 | -1.769 |
| | p | 0.070 | 0.782 | **0.045** | **0.034** | 0.780 | 0.444 | 0.318 | 0.168 | **0.039** | 0.193 | 0.133 | 0.079 |
| Actual age | b | -0.046 | -0.026 | -0.058 | -0.005 | -0.005 | -0.035 | 0.013 | 0.007 | -0.100 | -0.065 | -0.114 | -0.115 |
| | t | -3.787 | -1.232 | -3.830 | -0.258 | -0.262 | -1.054 | 0.595 | 0.247 | -7.516 | -2.853 | -6.880 | -4.647 |
| | p | **0.000** | 0.220 | **0.000** | 0.797 | 0.793 | 0.294 | 0.552 | 0.805 | **0.000** | **0.005** | **0.000** | **0.000** |
| Duration | b | -0.017 | -0.005 | -0.027 | -0.103 | 0.045 | -0.029 | 0.105 | 0.120 | 0.039 | 0.027 | 0.047 | 0.044 |
| | t | -0.922 | -0.176 | -1.123 | -2.901 | 1.668 | -0.671 | 3.090 | 2.315 | 1.978 | 0.936 | 1.781 | 1.046 |
| | p | 0.357 | 0.861 | 0.262 | **0.004** | 0.096 | 0.504 | **0.002** | **0.022** | **0.049** | 0.351 | 0.076 | 0.298 |
| Education | b | 0.083 | 0.056 | 0.098 | 0.084 | 0.011 | 0.069 | -0.029 | -0.006 | 0.147 | 0.239 | 0.111 | 0.061 |
| | t | 2.052 | 0.788 | 1.957 | 1.153 | 0.178 | 0.617 | -0.406 | -0.056 | 3.317 | 3.151 | 2.022 | 0.704 |
| | p | **0.041** | 0.432 | 0.052 | 0.251 | 0.859 | 0.539 | 0.685 | 0.956 | **0.001** | **0.002** | **0.044** | 0.482 |
| Sex | b | 0.436 | - | - | - | -0.185 | - | - | - | 0.091 | - | - | - |
| | t | 1.758 | - | - | - | -0.502 | - | - | - | 0.337 | - | - | - |
| | p | 0.080 | - | - | - | 0.616 | - | - | - | 0.736 | - | - | - |
| UPDRS III | b | -0.017 | -0.027 | -0.009 | -0.020 | 0.056 | 0.068 | 0.046 | 0.028 | -0.009 | 0.005 | -0.012 | -0.023 |
| | t | -1.853 | -1.713 | -0.825 | -1.256 | 4.163 | 2.757 | 2.851 | 1.242 | -0.916 | 0.321 | -0.936 | -1.243 |
| | p | 0.065 | 0.089 | 0.410 | 0.212 | **0.000** | **0.007** | **0.005** | 0.217 | 0.360 | 0.749 | 0.350 | 0.216 |

PD: Parkinson's disease; F: females; M: males; M*: matched males; MoCA: Montreal Cognitive Assessment; REM: Sleep Behavior Disorder Questionnaire Score; BJLO: Benton Judgment of Line Orientation Score; HVLT: Hopkins Verbal Learning Test. Significant results are highlighted in bold.

Table 5: Regression Model Outputs in PD: Cognitive Symptoms (Continued)

| Model | | Sleep Behavior Disorder Questionnaire Score (REM) | | | | Benton Judgement of Line Orientation Score (BJLO) | | | | HVLT Delayed Recall | | | |
|---|---|---|---|---|---|---|---|---|---|---|---|---|---|
| | | PD | PD-F | PD-M | PD-M* | PD | PD-F | PD-M | PD-M* | PD | PD-F | PD-M | PD-M* |
| Brain-PAD | b | 0.057 | 0.027 | 0.072 | 0.119 | -0.034 | -0.041 | -0.030 | -0.061 | -0.002 | 0.007 | -0.006 | -0.016 |
| | t | 3.097 | 0.884 | 3.153 | 3.536 | -2.365 | -1.570 | -1.764 | -2.175 | -0.107 | 0.247 | -0.298 | -0.490 |
| | p | **0.002** | 0.378 | **0.002** | **0.001** | **0.019** | 0.119 | 0.079 | **0.032** | 0.915 | 0.805 | 0.766 | 0.625 |
| Actual age | b | -0.018 | -0.043 | -0.007 | -0.024 | -0.032 | -0.047 | -0.024 | -0.010 | -0.080 | -0.076 | -0.083 | -0.067 |
| | t | -1.358 | -1.947 | -0.414 | -1.072 | -3.047 | -2.495 | -1.818 | -0.540 | -6.502 | -3.854 | -5.198 | -3.054 |
| | p | 0.175 | 0.054 | 0.679 | 0.286 | **0.002** | **0.014** | 0.070 | 0.590 | **0.000** | **0.000** | **0.000** | **0.003** |
| Duration | b | -0.006 | -0.063 | 0.040 | 0.095 | -0.002 | -0.011 | 0.005 | 0.000 | 0.004 | 0.018 | -0.007 | -0.027 |
| | t | -0.307 | -2.227 | 1.465 | 2.422 | -0.147 | -0.438 | 0.233 | 0.001 | 0.230 | 0.700 | -0.281 | -0.717 |
| | p | 0.759 | **0.028** | 0.144 | **0.017** | 0.883 | 0.662 | 0.816 | 0.999 | 0.818 | 0.485 | 0.779 | 0.474 |
| Education | b | -0.017 | -0.045 | -0.015 | -0.020 | 0.170 | 0.166 | 0.171 | 0.188 | 0.200 | 0.165 | 0.218 | 0.245 |
| | t | -0.378 | -0.610 | -0.260 | -0.254 | 4.839 | 2.635 | 3.967 | 2.822 | 4.880 | 2.505 | 4.127 | 3.176 |
| | p | 0.705 | 0.543 | 0.795 | 0.800 | **0.000** | **0.010** | **0.000** | **0.006** | **0.000** | **0.014** | **0.000** | **0.002** |
| Sex | b | -0.242 | - | - | - | -0.994 | - | - | - | 1.003 | - | - | - |
| | t | -0.873 | - | - | - | -4.602 | - | - | - | 3.993 | - | - | - |
| | p | 0.383 | - | - | - | **0.000** | - | - | - | **0.000** | - | - | - |
| UPDRS-III | b | 0.060 | 0.060 | 0.056 | 0.056 | -0.017 | -0.012 | -0.021 | -0.040 | -0.009 | -0.015 | -0.007 | -0.004 |
| | t | 5.953 | 3.715 | 4.319 | 3.225 | -2.119 | -0.833 | -2.174 | -2.795 | -1.021 | -1.001 | -0.543 | -0.231 |
| | p | **0.000** | **0.000** | **0.000** | **0.002** | **0.035** | 0.407 | **0.031** | **0.006** | 0.308 | 0.319 | 0.587 | 0.818 |

PD: Parkinson's disease; F: females; M: males; M*: matched males; MoCA: Montreal Cognitive Assessment; REM: Sleep Behavior Disorder Questionnaire Score; BJLO: Benton Judgment of Line Orientation Score; HVLT: Hopkins Verbal Learning Test. Significant results are highlighted in bold.

Table 5: Regression Model Outputs in PD: Cognitive Symptoms (Continued).

| Model | | HVLT Delayed Recognition | | | | olfactory testing | | | | Symbol Digit Modalities Score | | | |
|---|---|---|---|---|---|---|---|---|---|---|---|---|---|
| | | PD | PD-F | PD-M | PD-M* | PD | PD-F | PD-M | PD-M* | PD | PD-F | PD-M | PD-M* |
| Brain-PAD | b | 0.001 | -0.005 | 0.003 | -0.006 | -0.047 | 0.034 | -0.086 | 0.024 | -0.092 | -0.181 | -0.044 | -0.055 |
| | t | 0.077 | -0.404 | 0.310 | -0.438 | -0.820 | 0.326 | -1.236 | 0.204 | -1.498 | -1.624 | -0.598 | -0.517 |
| | p | 0.939 | 0.687 | 0.757 | 0.662 | 0.413 | 0.745 | 0.218 | 0.839 | 0.135 | 0.107 | 0.550 | 0.606 |
| Actual age | b | -0.029 | -0.021 | -0.034 | -0.023 | -0.266 | -0.279 | -0.253 | -0.279 | -0.398 | -0.396 | -0.386 | -0.340 |
| | t | -5.233 | -2.370 | -4.657 | -2.562 | -6.285 | -3.713 | -4.829 | -3.543 | -8.823 | -4.948 | -6.947 | -4.729 |
| | p | **0.000** | **0.019** | **0.000** | **0.012** | **0.000** | **0.000** | **0.000** | **0.001** | **0.000** | **0.000** | **0.000** | **0.000** |
| Duration | b | 0.008 | 0.023 | -0.005 | -0.012 | 0.057 | 0.078 | 0.041 | -0.038 | 0.002 | 0.020 | -0.015 | -0.086 |
| | t | 0.924 | 2.059 | -0.426 | -0.785 | 0.905 | 0.802 | 0.493 | -0.282 | 0.036 | 0.192 | -0.164 | -0.700 |
| | p | 0.356 | **0.042** | 0.671 | 0.434 | 0.366 | 0.424 | 0.623 | 0.779 | 0.972 | 0.848 | 0.870 | 0.486 |
| Education | b | 0.062 | 0.059 | 0.068 | 0.131 | 0.077 | 0.190 | 0.020 | -0.014 | 0.713 | 0.666 | 0.781 | 0.861 |
| | t | 3.354 | 2.033 | 2.816 | 4.203 | 0.548 | 0.760 | 0.116 | -0.050 | 4.751 | 2.502 | 4.259 | 3.420 |
| | p | **0.001** | **0.044** | **0.005** | **0.000** | 0.584 | 0.449 | 0.907 | 0.960 | **0.000** | **0.014** | **0.000** | **0.001** |
| Sex | b | 0.256 | - | - | - | 2.392 | - | - | - | 3.159 | - | - | - |
| | t | 2.251 | - | - | - | 2.770 | - | - | - | 3.434 | - | - | - |
| | p | **0.025** | - | - | - | **0.006** | - | - | - | **0.001** | - | - | - |
| UPDRS-III | b | -0.010 | -0.010 | -0.009 | -0.010 | -0.055 | -0.047 | -0.057 | -0.062 | -0.145 | -0.059 | -0.194 | -0.208 |
| | t | -2.364 | -1.551 | -1.610 | -1.537 | -1.736 | -0.845 | -1.444 | -1.039 | -4.308 | -1.000 | -4.629 | -3.820 |
| | p | **0.019** | 0.124 | 0.109 | 0.127 | 0.083 | 0.400 | 0.150 | 0.301 | **0.000** | 0.319 | **0.000** | **0.000** |

PD: Parkinson's disease; F: females; M: males; M*: matched males; MoCA: Montreal Cognitive Assessment; REM: Sleep Behavior Disorder Questionnaire Score; BJLO: Benton Judgment of Line Orientation Score; HVLT: Hopkins Verbal Learning Test. Significant results are highlighted in bold.

### 3.5. Brain age, mood, and anxiety

While brain-PAD was not associated with anxiety or GDS scores, these symptom scores were associated with increased MoCA scores in all groups (p<0.001; Table 6). Anxiety scores were negatively associated with Actual age in PD (t(364)= -3.79, p < 0.001), PD-M (t(237) = -3.76, p < 0.001), and PD-M* (t(122) = -2.12, p = 0.036), but not in PD-F (t(121) = -1.38, p = 0.17). In addition, anxiety was associated with the duration of the disease (t(364) = -2.09, p = 0.037) and level of education (t(364) = -2.51, p = 0.012) in the PD group.

Table 6: Regression Model Outputs in PD: Mood and Anxiety

| Model | | Anxiety | | | | GDS | | | |
|---|---|---|---|---|---|---|---|---|---|
| | | PD | PD-F | PD-M | PD-M* | PD | PD-F | PD-M | PD-M* |
| Brain-PAD | b | 0.132 | 0.282 | 0.063 | -0.054 | 0.004 | -0.026 | 0.018 | 0.000 |
| | t | 1.053 | 1.235 | 0.422 | -0.221 | 0.225 | -0.876 | 0.902 | -0.008 |
| | p | 0.293 | 0.219 | 0.673 | 0.825 | 0.822 | 0.383 | 0.368 | 0.994 |
| Actual age | b | -0.356 | -0.228 | -0.437 | -0.343 | -0.030 | -0.040 | -0.024 | -0.012 |
| | t | -3.791 | -1.381 | -3.765 | -2.122 | -2.418 | -1.884 | -1.555 | -0.568 |
| | p | **0.000** | 0.170 | **0.000** | **0.036** | **0.016** | 0.062 | 0.121 | 0.571 |
| Duration | b | -0.286 | -0.607 | -0.051 | 0.074 | -0.005 | -0.052 | 0.031 | 0.060 |
| | t | -2.091 | -2.873 | -0.284 | 0.260 | -0.291 | -1.892 | 1.281 | 1.562 |
| | p | **0.037** | **0.005** | 0.777 | 0.796 | 0.771 | 0.061 | 0.201 | 0.121 |
| Education | b | -0.776 | -1.080 | -0.616 | -0.492 | -0.100 | -0.099 | -0.102 | -0.047 |
| | t | -2.517 | -1.977 | -1.642 | -0.866 | -2.443 | -1.412 | -2.016 | -0.611 |
| | p | **0.012** | 0.050 | 0.102 | 0.388 | **0.015** | 0.161 | **0.045** | 0.542 |
| Sex | b | 2.899 | - | - | - | -0.135 | - | - | - |
| | t | 1.536 | - | - | - | -0.538 | - | - | - |
| | p | 0.125 | - | - | - | 0.591 | - | - | - |
| MoCA | b | 0.390 | 0.413 | 0.389 | 0.440 | 0.057 | 0.072 | 0.048 | 0.043 |
| | t | 5.643 | 3.377 | 4.568 | 3.571 | 6.236 | 4.572 | 4.195 | 2.607 |
| | p | **0.000** | **0.001** | **0.000** | **0.001** | **0.000** | **0.000** | **0.000** | **0.010** |
| UPDRS-III | b | -0.333 | 0.492 | -0.778 | -0.706 | 0.000 | 0.090 | -0.031 | -0.141 |
| | t | -0.840 | 0.705 | -1.614 | -1.003 | -0.004 | 0.999 | -0.472 | -1.499 |
| | p | 0.402 | 0.482 | 0.108 | 0.318 | 0.997 | 0.320 | 0.637 | 0.136 |

PD: Parkinson's disease; F: females; M: males; M*: matched males; GDS: Geriatric Depression Scale. Significant results are highlighted in bold.

### 3.6 CSF Biomarkers and Other Variables in PD

Interestingly, a-syn level was associated with brain-PAD, Actual age, and duration of disease in PD (all) and PD-F groups, but not in PD-M or PD-M* (Table 7). Amyloid-beta was also associated with brain-PAD in the PD (t(354) = -3.28, p = 0.001), PD-F (t(118) = -2.89, p = 0.005), and PD-M (t(232) = -2.10, p = 0.046) groups, while actual age was not correlated with amyloid-beta level. The CSF p-tau level was not correlated with Brain-PAD, but it was correlated with actual age in all subgroups (p<0.036) (Table 7). Lower caudate measures were found to be associated with higher actual age in PD (t(364) = -3.89, p < 0.001), PD-F (t(124) = -2.11, p = 0.036), and PD-M (t(236) = -3.22, p = 0.001). In the PD-M* group, an association was only found between caudate measures and brain-PAD (t(121) = -2.01, P = 0.016). Putamen measures were also significantly correlated with actual age in PD (t(364)=-2.332, p=0.020) and PD-M (t(236)=-2.321, p=0.021), but it was not replicated in PD-F nor PD-M*(Table 8).

Table 7: Regression Model Outputs in PD: CSF Biomarkers (pg/mL)

| Model | | a-Synuclein (a-syn) | | | | Amyloid-b | | | | CSF p-tau (2016 assay) | | | |
|---|---|---|---|---|---|---|---|---|---|---|---|---|---|
| | | PD | PD-F | PD-M | PD-M* | PD | PD-F | PD-M | PD-M* | PD | PD-F | PD-M | PD-M* |
| Brain-PAD | b | -15.45 | -32.68 | -8.89 | -10.69 | -10.39 | -19.479 | -6.897 | -7.799 | -0.050 | -0.160 | -0.015 | 0.049 |
| | t | -3.117 | -3.644 | -1.517 | -1.251 | -3.280 | -2.890 | -2.010 | -1.505 | -1.206 | -1.753 | -0.328 | 0.707 |
| | p | **0.002** | **0.000** | 0.131 | 0.213 | **0.001** | **0.005** | **0.046** | 0.135 | 0.229 | 0.082 | 0.743 | 0.481 |
| Actual age | b | 10.51 | 15.43 | 8.042 | 5.669 | 0.577 | 0.849 | 0.651 | -0.097 | 0.135 | 0.168 | 0.122 | 0.101 |
| | t | 2.988 | 2.624 | 1.866 | 0.977 | 0.256 | 0.190 | 0.259 | -0.028 | 4.684 | 2.893 | 3.837 | 2.126 |
| | p | **0.003** | **0.010** | 0.063 | 0.331 | 0.798 | 0.850 | 0.796 | 0.978 | **0.000** | **0.005** | **0.000** | **0.035** |
| Duration | b | 12.13 | 27.21 | 0.624 | 11.672 | 0.944 | 9.153 | -4.865 | -1.162 | 0.024 | 0.106 | -0.047 | 0.058 |
| | t | 2.330 | 3.580 | 0.090 | 1.238 | 0.283 | 1.583 | -1.204 | -0.203 | 0.565 | 1.501 | -0.903 | 0.523 |
| | p | **0.020** | **0.000** | 0.928 | 0.218 | 0.777 | 0.116 | 0.230 | 0.839 | 0.573 | 0.136 | 0.368 | 0.601 |
| Education | b | -3.129 | 4.472 | -6.392 | -13.122 | -4.139 | 5.271 | -9.018 | -5.135 | 0.067 | 0.050 | 0.080 | 0.110 |
| | t | -0.262 | 0.227 | -0.433 | -0.697 | -0.544 | 0.355 | -1.051 | -0.452 | 0.696 | 0.269 | 0.736 | 0.676 |
| | p | 0.794 | 0.821 | 0.666 | 0.487 | 0.587 | 0.723 | 0.294 | 0.652 | 0.487 | 0.789 | 0.462 | 0.500 |
| Sex | b | 106.4 | - | - | - | 26.29 | - | - | - | 0.543 | - | - | - |
| | t | 1.466 | - | - | - | 0.564 | - | - | - | 0.922 | - | - | - |
| | p | 0.144 | - | - | - | 0.573 | - | - | - | 0.357 | - | - | - |

PD: Parkinson's disease; F: females; M: males; M*: matched males. Significant results are highlighted in bold.

Table 8: Regression Model Outputs in PD: Other Variables.

| Model | | Caudate (L+R) | | | | Putamen (L+R) | | | |
|---|---|---|---|---|---|---|---|---|---|
| | | PD | PD-F | PD-M | PD-M* | PD | PD-F | PD-M | PD-M* |
| Brain-PAD | b | -0.008 | 0.008 | -0.015 | -0.031 | -0.004 | 0.006 | -0.008 | -0.006 |
| | t | -0.956 | 0.543 | -1.603 | -2.013 | -0.834 | 0.741 | -1.587 | -0.814 |
| | p | 0.340 | 0.588 | 0.110 | **0.016** | 0.405 | 0.460 | 0.114 | 0.417 |
| Actual age | b | -0.023 | -0.023 | -0.022 | -0.018 | -0.007 | -0.004 | -0.008 | -0.005 |
| | t | -3.892 | -2.117 | -3.221 | -1.786 | -2.332 | -0.721 | -2.321 | -0.946 |
| | p | **0.000** | **0.036** | **0.001** | 0.077 | **0.020** | 0.472 | **0.021** | 0.346 |
| Duration | b | 0.001 | 0.000 | 0.000 | 0.005 | -0.010 | -0.014 | -0.006 | -0.003 |
| | t | 0.066 | 0.034 | 0.013 | 0.268 | -2.126 | -1.934 | -1.040 | -0.295 |
| | p | 0.947 | 0.973 | 0.990 | 0.789 | 0.034 | 0.055 | 0.299 | 0.769 |
| Education | b | 0.010 | -0.003 | 0.017 | 0.008 | -0.013 | -0.013 | -0.014 | -0.017 |
| | t | 0.526 | -0.079 | 0.726 | 0.229 | -1.264 | -0.641 | -1.116 | -0.982 |
| | p | 0.600 | 0.937 | 0.469 | 0.819 | 0.207 | 0.523 | 0.265 | 0.328 |
| Sex | b | 0.116 | - | - | - | 0.035 | - | - | - |
| | t | 0.956 | - | - | - | 0.544 | - | - | - |
| | p | 0.340 | - | - | - | 0.587 | - | - | - |

PD: Parkinson's disease; F: females; M: males; M*: matched males. Significant results are highlighted in bold.

### 3.7 Association between GM/WM changes and Brain-PAD

To visualize the brain regions that significantly contributed to brain-PAD score estimation, the brain-PAD scores were regressed to GM/WM tissue probability maps in VBM analysis. As expected, negative correlations with GM/WM volumes were observed throughout the brain (Tables 9 and 10, and Fig. 3). In the healthy group, multiple regression analysis revealed a significant decrease in GM volume with increasing brain-PAD scores in the parietal lobe, limbic lobe, middle temporal gyrus, hippocampus, parahippocampal gyrus, and frontal lobe regions. In addition, healthy subjects showed decreased WM volume within the frontal lobe and limbic lobe regions. In the PD group, we observed a significant reduction in GM volume located extensively in the primary frontal, temporal, parietal, limbic, and occipital lobes. Higher brain-PAD scores in the PD group were associated with decreased WM volumes in the cerebellum, lentiform nucleus, medulla, midbrain, and frontal lobe. There was no positive association between brain-PAD scores and GM/WM volumes in either the HC or PD groups.

Table 9: Clusters of negative grey matter and white matter alterations with brain-PAD in healthy controls (N = 105).

| Analysis | Cluster | Region | BA | Cluster size (no. of voxels) | P (FWE) # | Hemisphere | MNI coordinats (X, y, z) | T value (peak voxel) |
|---|---|---|---|---|---|---|---|---|
| GM | 1 | Parietal Lobe/ Supramarginal Gyrus | 40 | 399 | 0.000 | L | -66, -46, 21 | 6.30 |
|  | 2 | Limbic Lobe/ Cingulate_Mid | 24/31 | 385 | 0.000 | R | 3, -9, 45 | 6.13 |
|  |  |  |  |  |  | R | 6, -22, 42 | 5.27 |
|  |  |  |  |  |  | R | 8, -14, 38 | 4.95 |
|  | 3 | Middle Temporal Gyrus/ Temporal_Inf | 21/38 | 567 | 0.001 | R | 51, 4, -33 | 6.02 |
|  |  |  |  |  |  | R | 57, -6, -26 | 4.92 |
|  | 4 | Hippocampus/ Limbic Lobe/ Parahippocampa Gyrus | - | 234 | 0.002 | L | -27, -15, -14 | 5.68 |
|  | 5 | Frontal Lobe/ Frontal_Inf_Tri | - | 108 | 0.005 | L | -42, 45, 8 | 5.48 |
|  | 6 | Fusiform/ Limbic Lobe/ Parahippocampa Gyrus | 19/37 | 207 | 0.005 | R | 24, -52, -12 | 5.45 |
|  |  |  |  |  |  | R | 22, -44, -12 | 5.29 |
| WM | 1 | Frontal Lobe/ Limbic Lobe/ Sub-Gyral | - | 317 | 0.002 | L | -16, 27, 24 | 5.63 |

BA = Brodmann Area; R = right hemisphere; L = left hemisphere; GM = gray matter; WM = white matter; HC = healthy controls; MNI = Montreal Neurological Institute; FWE = family-wise error false discovery rate; # = the p-value reported at peak-level based on FWE corrections.

Table 10: Clusters of negative grey matter and white matter alterations with Brain-PAD in Parkinson's disease (N = 373)

| Analysis | Cluster | Region | BA | Cluster size (no. of voxels) | P# (FWE) | Hemisphere | MNI coordinats (X, y, z) | T value (peak voxel) |
|---|---|---|---|---|---|---|---|---|
| GM | 1 | Frontal Lobe/ Temporal Lobe/ Parietal Lobe/ Limbic Lobe | 40/32/13/31/10/22/21/ | 100439 | 0.000 | L<br>L | -2, 27, 30<br>-4, -64, 15 | 9.39<br>8.87 |
| | 2 | Frontal Lobe/ Middle Frontal Gyrus/ Superior Frontal Gyrus | 10/9/46/8 | 5922 | 0.000 | R<br>R | 40,40, 32<br>48,40,14 | 8.00<br>7.54 |
| | 3 | Frontal Lobe/ Superior Frontal Gyrus | 11 | 175 | 0.000 | L | -28, 45, -21 | 5.86 |
| | 4 | Frontal Lobe/ Inferior Frontal Gyrus/ Superior Frontal Gyrus | 11 | 723 | 0.000 | R<br>R | 26, 36, -22<br>27, 46, -22 | 5.81<br>5.73 |
| | 5 | Occipital Lobe/ Parietal Lobe/ Precuneus | 19 | 421 | 0.000 | R<br>R | 32, -82, 38<br>34, -90, 24 | 5.79<br>5.62 |
| | 6 | Frontal Lobe/ Middle Frontal Gyrus | 10 | 394 | 0.000 | L<br>L | -50, 48, 0<br>-46, 54, -6 | 5.71<br>5.70 |
| WM | 1 | Sub-lobar/ Midbrain/ Lentiform Nucleus | - | 1426 | 0.000 | L | -16, -9, -2 | 6.24 |
| | 2 | Cerebellum/ Medulla/ Left Brainstem | - | 549 | 0.000 | L<br>R | -9, -40, -60<br>9, -40, -60 | 5.61<br>4.80 |
| | 3 | Cerebrum/Sub-lobar/ Lentiform Nucleus | - | 365 | 0.002 | R | 20, -10, 6 | 5.20 |
| | 4 | Frontal Lobe/ Middle Frontal Gyrus | 47 | 181 | 0.004 | R | 20, 33, -14 | 4.98 |
| | 5 | Cerebrum/ Frontal Lobe/ Anterior Cingulate/ Limbic Lobe | - | 208 | 0.008 | L | -16, 30, 21 | 4.84 |
| | 6 | Rectus/ Medial Frontal Gyrus | 47 | 109 | 0.014 | L | -12, 28, -15 | 4.70 |

BA = Brodmann Area; R = right hemisphere; L = left hemisphere; GM = gray matter; WM = white matter; PD = Parkinson's disease; MNI = Montreal Neurological Institute; FWE = family-wise error false discovery rate; # = the p-value reported at peak-level based on FWE corrections.

**Fig. 3**. Negative association between gray and white matter volumes and brain-PAD scores in healthy control and PD groups (corrected using FW at p ≤ .05 and an extent threshold of K = 100).

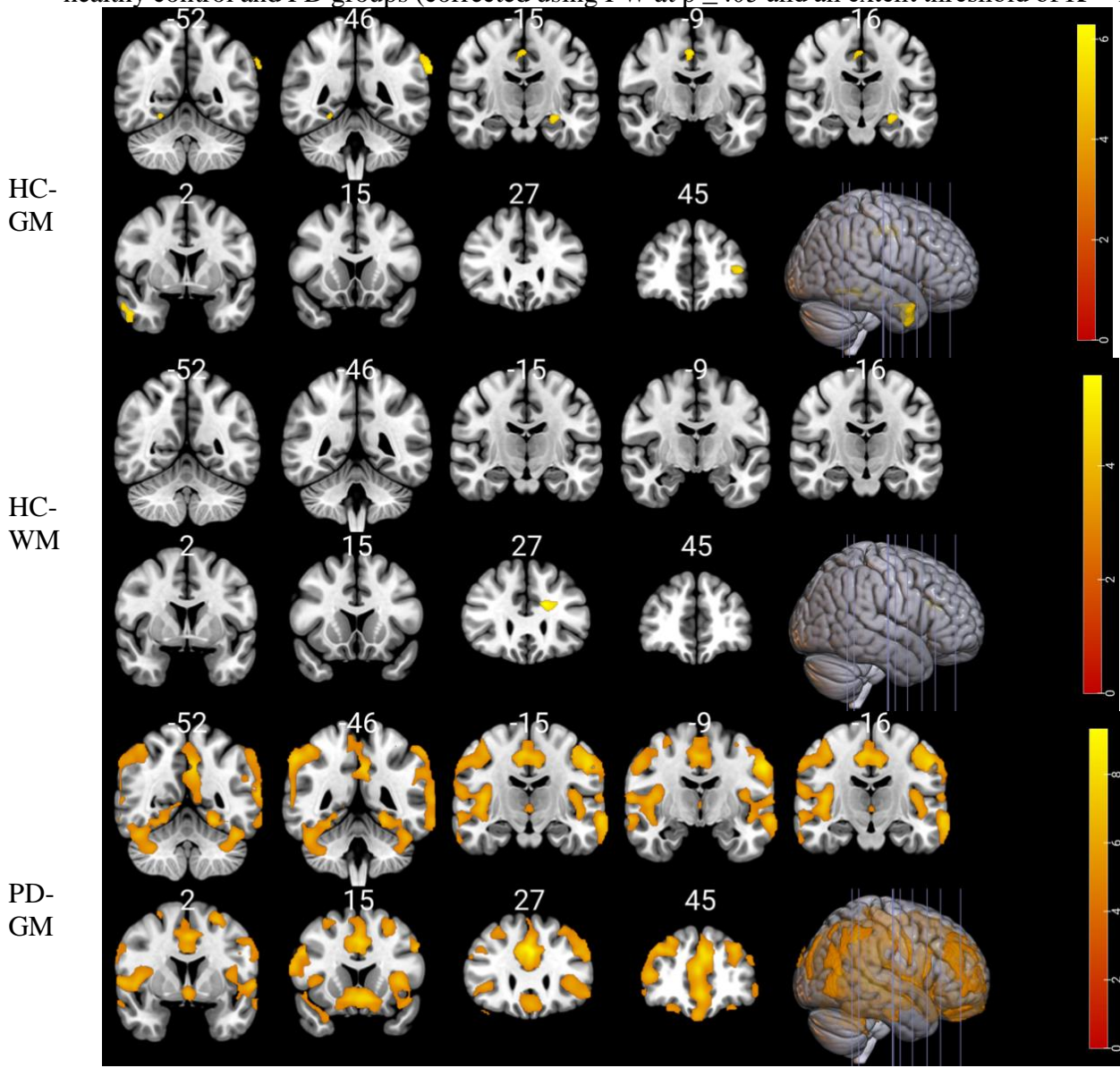

PD-
WM

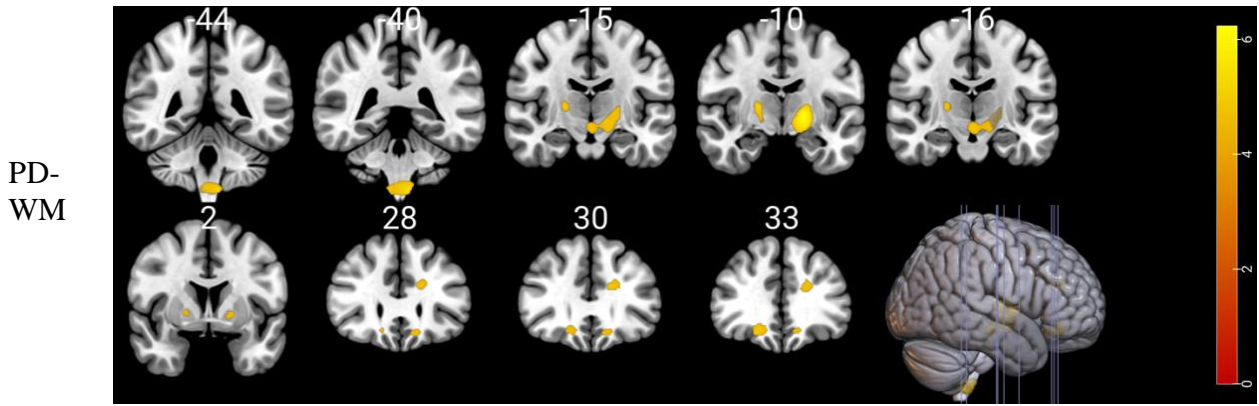

Figure 4 illustrates the regions where the regression slopes between the HC and PD groups diverged in GM, as well as the correlation between GM volumes and Brain-PAD in each group. A significant cluster, consisting of 1200 voxels, was predominantly located in the left Parahippocampal Gyrus, extending to Hippocampus and Amygdala. Notably, the HC group exhibited a significant negative correlation between GM volumes and Brain-PAD in this region ($r = -0.44$, $p < 0.001$), whereas the PD group did not demonstrate such a correlation (Fig. 4B; $r = -0.05$, $p < 0.33$). There was no significant difference in the regression slopes between the HC and PD groups in WM.

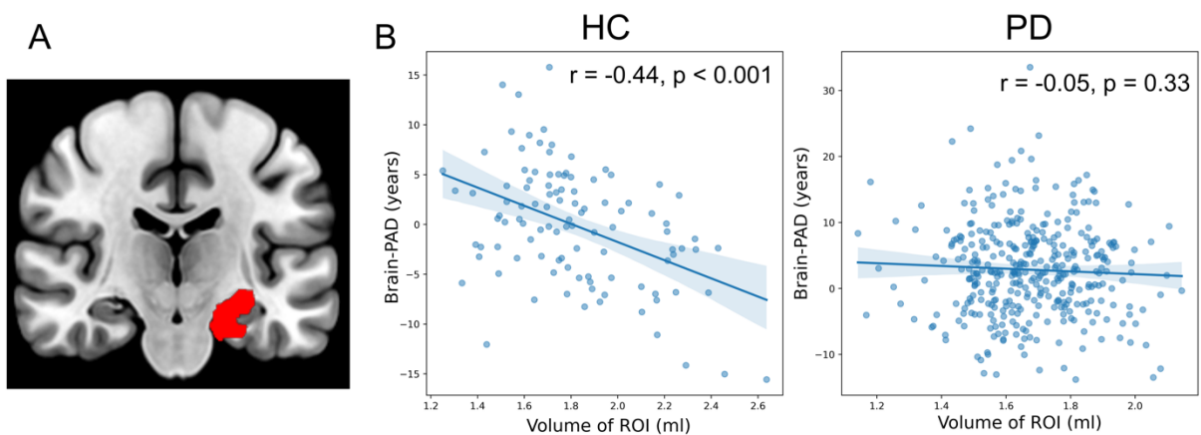

**Fig. 4.** (A): The brain region showing a significant difference in regression slopes between the HC and PD groups in GM. (B): The correlation between GM volumes in the region that was identified to be significant and Brain-PAD within each group.

Figure 5 illustrates the association between volumes of GM/WM and Brain-PAD scores within each sex (i.e., PD-F, PD-M, PD-M*). Notably, we observed a significant reduction in GM volume across all subgroups as Brain-PAD scores increased (Tables 11 and Figure 5). In the PD-M group, similar topology of negative correlation between Brain-PAD scores and GM volumes was observed as identified the regression analysis using the whole PD patients (Fig 3), while lesser degree of association was observed in PD-M* and PD-F, potentially due to the smaller sample size. No negative correlation was observed between Brain-PAD scores and WM volumes in all PD subgroups. Similarly, no positive association was observed between Brain-PAD scores and GM/WM volumes in the PD subgroups. The difference in regression slopes between the PD-F and PD-M* groups was not statistically significant for either GM or WM modalities.

Table 11: Clusters of negative grey matter alterations with Brain-PAD in in PD group with respect to the sex categories (PD-F, N = 129; PD-M, N = 244; PD-M*, N = 129).

| Analysis | Cluster | Region | BA | Cluster size (no. of voxels) | P# (FWE) | Hemisphere | MNI coordinats (X, y, z) | T value (peak voxel) |
|---|---|---|---|---|---|---|---|---|
| PD-F | 1 | Frontal Lobe/ Anterior Cingulate/Limbic Lobe | 25 | 789 | 0.000 | L | 0, 4, -14 | 6.73 |
| | 2 | Temporal Lobe/ Middle Temporal Gyrus | 22 | 245 | 0.000 | R R | 62, -36, 3 48,-32,10 | 6.02 5.29 |
| | 3 | Frontal Lobe/ Inferior Frontal Gyrus/Insula | - | 258 | 0.000 | L | -39, 21, -2 | 5.86 |
| | 4 | Frontal Lobe/ Inferior Frontal Gyrus | 11 | 340 | 0.000 | L | -15, 28, -22 | 5.59 |
| | 5 | Cingulate Gyrus/ Limbic Lobe | 32 | 110 | 0.000 | R | 3, 9, 44 | 5.35 |
| | 1 | Temporal Lobe/ Parietal Lobe/Frontal Lobe/Postcentral Gyrus | 40/21/22 | 15552 | 0.000 | L L L | -63, -38, 21 -63, -12, -10 -66, -38, 3 | 7.93 7.88 7.44 |

| Group | # | Region | BA | Cluster size | p# | Hemisphere | MNI coordinates | Z-score |
|---|---|---|---|---|---|---|---|---|
| PD-M | 2 | Limbic Lobe/ Cingulate Gyrus/ Frontal Lobe/ Precuneus | 31/32/24 | 16830 | 0.000 | R | 26, -33, -18 | 7.61 |
| | | | | | | R | 3, 26, 34 | 7.48 |
| | | | | | | ? | 0, 46, 14 | 7.38 |
| | 3 | Parietal Lobe/Precentral Gyrus/Inferior Parietal Lobule | 4/40 | 5491 | 0.000 | R | 57, -46, 44 | 6.83 |
| | | | | | | R | 42, -20, 2 | 6.28 |
| | | | | | | R | 39, -20, 9 | 6.23 |
| | 4 | Fusiform/Limbic Lobe/Parahippocampa Gyrus | 36/19 | 1285 | 0.000 | L | -24, -39, -14 | 6.58 |
| | | | | | | L | -34, -34, -22 | 5.45 |
| | | | | | | L | -27, -63, -12 | 5.43 |
| | 5 | Frontal Lobe/ Middle Frontal Gyrus/Superior Frontal Gyrus | 9/46/10 | 1844 | 0.000 | L | -32, 42, 33 | 6.57 |
| | | | | | | L | -44, 30, 28 | 6.30 |
| | | | | | | L | -30, 51, 21 | 5.84 |
| PD-M* | 1 | Temporal Lobe/ Middle Temporal Gyrus | 21 | 345 | 0.000 | L | -62, -10, -12 | 6.66 |
| | 2 | Frontal Lobe/ Medial Frontal Gyrus | 9 | 241 | 0.000 | L | -4, -42, 28 | 6.18 |
| | 3 | Cerebrum/Superior Temporal Gyrus/Temporal Lobe | - | 140 | 0.001 | L | -57, 0, 2 | 5.77 |
| | 4 | Frontal Lobe/ Subcallosal Gyrus/Inferior Frontal Gyrus | 47 | 162 | 0.001 | L | -14, 12, -15 | 5.68 |
| | 5 | Cerebrum/ Frontal Lobe/ Superior Temporal Gyrus | 22 | 259 | 0.000 | R | 48, 4, 0 | 5.67 |
| | | | | | | R | 57, 4, 2 | 5.33 |

BA = Brodmann Area; R = right hemisphere; L = left hemisphere; GM = gray matter; WM = white matter; PD = Parkinson's disease; MNI = Montreal Neurological Institute; FWE = family-wise error false discovery rate; # = the p-value reported at peak-level based on FWE corrections.

**Fig. 5**. Negative association between gray matter volumes and brain-PAD scores in PD group within each sex (corrected using FWE at p ≤ .05 and an extent threshold of K = 100).

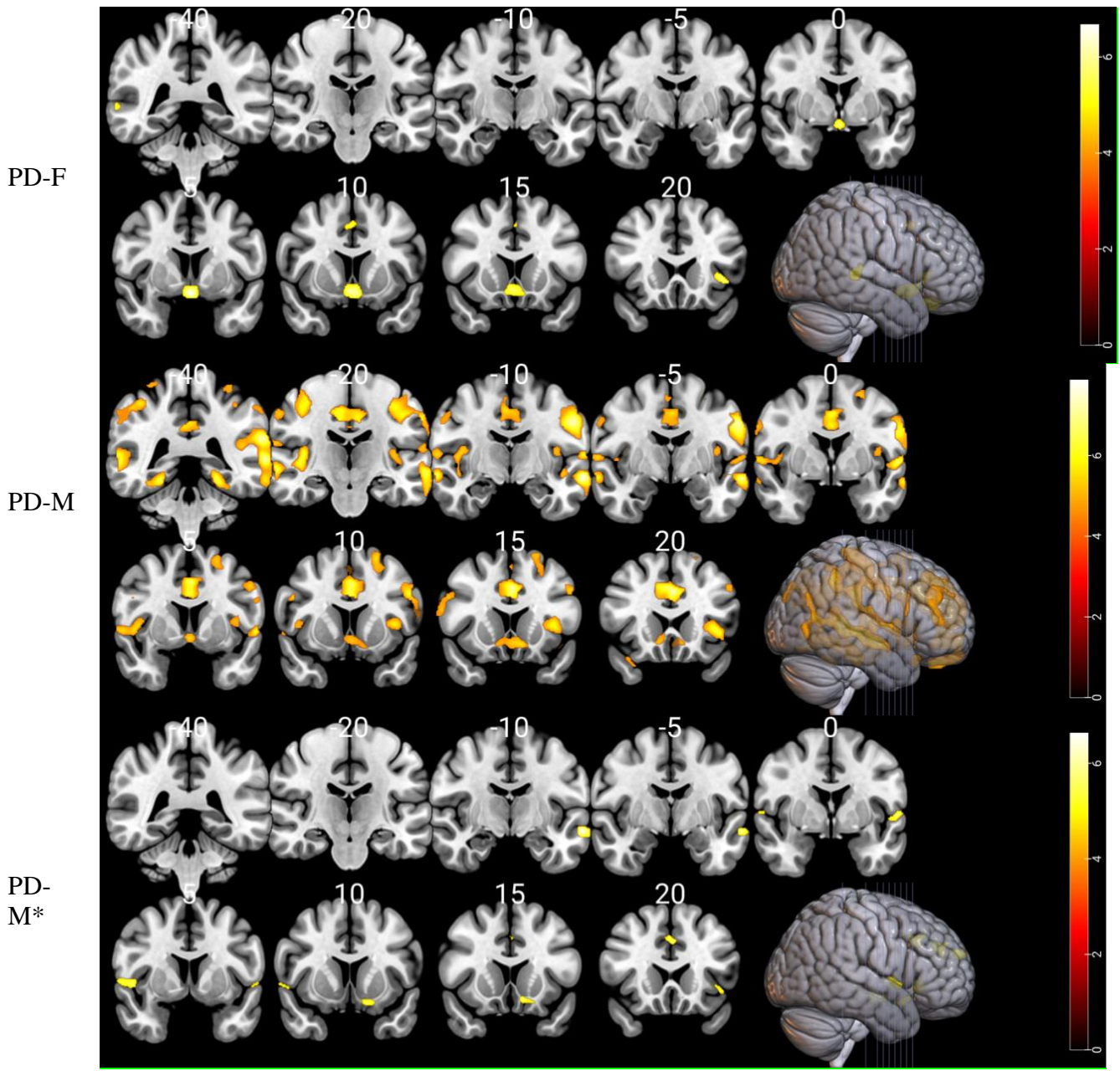

PD-F

PD-M

PD-M*

## 4. Discussion

In this study, we developed a brain age estimation model based on multi-site and multi-scanner datasets, such as IXI, OASIS, and PPMI, demonstrating the generalizability of our results. Of note, the MRI processing technique used in this study has been thoroughly examined and found to be suitable for multi-center and multi-scanner studies [20]. We then applied this methodology to investigate the sex-specific association between accelerated brain aging and PD pathology. We found that PD patients had a significantly greater Brain-PAD than healthy controls (t(476)= 3.94, p< 0.0001), which is in line with what has been observed in previous studies [15,16]. Additionally, our Brain-PAD was significantly greater in male PD patients compared to female patients (t(371)=2.26, p=0.024). This sex difference was still significant when propensity score-matched male patients were used (t(256) = 2.50, p = 0.012). Of note, our brain age prediction model did not show a significant difference in brain-PAD between male and female HC (training set: t(947) = 0.52, p = 0.60; hold-out set: t(103) = 0.56, p = 0.57). This result would suggest that male sex is more influenced by disease-accelerated brain atrophy associated with aging in PD [7,21].

Several previous studies have investigated sex differences in brain structure in PD using structural brain imaging, with some variability in the results [7]. In established PD, increased atrophy has been observed in males in extensive cortical and subcortical regions, as measured using VBM as well as cortical thickness [7,21]. According to a study conducted on the PPMI dataset, male de novo PD patients exhibit greater cortical thinning in the pre- and post-central gyrus and lower volume of the basal ganglia, thalamus, and brain-stem compared to their female counterparts [22]. Male sex has also been shown to be a significant predictor of greater atrophy and clinical severity in PD in the PPMI dataset [23]. While we did not directly compare brain structure between male and female patients in this study, we found that brain-PAD in the PD-M* was significantly higher than the PD-F group (Table 3). The clinical relevance of Brain-PAD in PD in general has been demonstrated by its significant correlation with the UPDRS-III score (Table 4), supporting the idea that brain age measurements can provide a clinically relevant metric for ascertaining the severity and extent of PD-induced neurodegeneration. As expected, UPDRS-III correlated with the patients' age and disease duration in this study cohort. While this correlation persisted when looking at just the PD-M group, the statistical significance was no longer observed when analyzed within only female PD patients (PD-F). The non-significance in PD-F may potentially be due to the lower sample size of female PD patients (n=244 vs. 129). For example, the correlation between Brain-PAD and UPDRS-III became non-significant when propensity score-matched male PD patients were used (PD-M*).

Our analysis showed key sex differences in the relationship between accelerated brain age and cognitive symptoms of PD. Interestingly, a lower MoCA score was significantly associated with Brain-PAD only in male PD (both PD-M and PD-M*, p<0.04) while it was not significant in female PD patients (p = 0.782). Additionally, visuospatial acuity (BJLO score) was negatively correlated with brain-PAD only in the pooled PD group and the PD-M* group, indicating this correlation is mainly driven by male patients. Male sex is significantly associated with higher incidence of MCI in de novo PD, as well as more rapid progression to dementia longitudinally [5,6,24]. More specifically, male PD patients exhibit impairments in overall cognition, processing speed, and working memory than females, whereas females tend to display more severe deficits in visuospatial function [6,25], although the relatively better visuospatial function in males is also observed in normal aging [26]. Deficit in visuospatial function in males, according to our data, is potentially tied to an accelerated brain aging phenotype which is expressed to a greater degree in males than in females and may partly underlie the observed sex differences in cognitive symptoms.

When looking at the association between brain-PAD and other symptoms of PD, our regression analysis revealed significant sex differences, particularly in non-motor symptoms of PD such as cognition and REM sleep behaviour disorder scores (RBD) (Table 5). The RBD score was significantly correlated with the brain-PAD score in our pooled cohort of PD patients. Additionally, when we looked at females with PD and propensity-matched males with PD separately, we found that the correlation between RBD and brain-PAD was significant only in the male group. RBD sleep disorder is one of the most common prodromal symptoms of PD and is a strong risk factor for the development of PD or other synucleinopathies [27]. Furthermore, the co-occurrence of RBD sleep disorder in individuals with PD is linked with inferior cognitive functioning, heightened levels of depression, and apathetic symptoms [27,28]. The presence of RBD in PD is often accompanied by a negative prognosis due to the correlation with autonomic dysfunction, heightened disease burden, and increased mortality rates as evidenced by various studies [29].

Very little is understood about the neuropathology of RBD, but it appears that the key neuronal networks in the brainstem that regulate skeletal muscle atonia during sleep are selectively vulnerable to synucleopathy [30]. While some studies have noted an increased incidence in male PD patients, others have found no apparent sex difference in the presentation of RBD [29]. Further investigation has shown a difference in the presentation of REM-sleep behaviour disorder, with male patients presenting with a significantly more violent form with more vigorous motor activity, which could possibly lead to a detection-bias for RBD in males [31]. A recent study found that male PD patients with RBD have significantly worse subcortical brain atrophy compared to female patients, and these sex differences are greater than those observed when comparing male and female patients without RBD [32]. Moreover, male patients with PD-RBD have severer cognitive symptoms. This, along with the results of our study, indicates that RBD coincides with greater macroscopic structural alterations in male patients compared to female patients.

To visualize the brain regions that mainly contributed to the brain-PAD score estimation, a regression model was constructed to examine the correlation between brain-PAD scores and volumes of GM and WM. Within the HC group (i.e., hold-out set), an elevation in brain-PAD demonstrated a significant association with decreased GM volume, particularly in the limbic area. Furthermore, an observed reduction in WM within the limbic region corresponded to increasing brain-PAD in the HC group, suggesting a potential link between elevated brain-PAD and cognitive decline as well as deficits in emotional processing among older HC individuals.

In the PD group, there was a notably greater magnitude of GM changes in response to increasing brain-PAD compared to WM. These findings suggest a stronger association between escalated brain-PAD and the degeneration of neuronal cell bodies, dendrites, and glial cells rather than myelinated nerve fibers (axons) among individuals with PD. Regarding the specific brain regions associated with increasing brain-PAD in PD, notable GM changes were primarily localized within the Frontal Lobe, Limbic Lobe, Superior Frontal Gyrus, and Brodmann Areas 31 and 10/46 (Dorsolateral Prefrontal Cortex). These regions are known to be involved in cognitive and executive functions. Additionally, the study demonstrated significant WM changes in response to greater brain-PAD in PD, primarily concentrated in the midbrain, basal ganglia, and cerebellum regions. These areas are particularly relevant to motor control, coordination, balance and posture, as well as the timing and coordination of movements (Fig 3).

In addition, our investigation included conducting regression analyses involving the interaction between GM/WM volume and Brain-PAD scores, with the aim of identifying dissimilarities between the HC and PD groups. We observed a significant distinction in the regression slopes between the HC and PD groups in relation to GM, primarily localized in the parahippocampal/hippocampal/amygdala area (Fig. 4 A). As expected, we found a significant correlation between GM volume in this ROI and Brain-PAD values within the HC group suggesting a significant contribution of this region's atrophy toward the overall brain age estimation (Fig. 4 B). However, this significant correlation was eliminated within the PD group, potentially suggesting that the disease-related degeneration interfered with the normal aging-related degeneration in this limbic area.

The cause of sex differences in PD symptomatology and incidence between male and female patients is not well understood, but may involve a neuroprotective effect of estrogens modulating neuroinflammation, metabolism, and signaling of the nigrostriatal dopamine system, along with other neurodevelopmental factors [33]. This is supported by the observation that the age of menopause, duration of fertile life, and levels of female sex hormones significantly affect the incidence and risk of PD in females [34,35]. Estrogens may also prevent aggregation and defibrillation of alpha-synuclein, reducing the deposition of Lewy bodies [36-38]. This mechanism could help explain the differences in regional atrophy and white matter integrity throughout the brain between sexes and would lead to the enhanced pattern of regional cortical atrophy among male PD patients, as reported in other studies[7]. These points could potentially explain why males with PD had faster brain aging than females in our study. It is worth mentioning that, despite the fact that male sex is associated with poorer prognosis, cognitive symptoms, and higher prevalence of PD, the converse holds true in the case of AD [39]. In the context of AD, women tend to perform worse on neurocognitive tests and exhibit a higher degree of atrophy, as compared to men [39,40]. This observation indicates that the higher incidence of disease and relatively worse outcomes for males in cognitive and motor symptoms in PD are not only due to underlying sex differences in healthy aging or a generalized increased vulnerability to neurodegeneration in males, but also an increased vulnerability specifically to synucleinopathies. Further longitudinal studies are necessary to accurately quantify the differences in brain aging between males and females with PD.

Our research furthers our insight into the distinctive neurological characteristics seen in male patients with PD. Our study has revealed a remarkable neurological presentation that appears as an accelerated aging-related phenotype in brain structure. Through examining the brain structure of male PD patients, we have noticed distinct patterns that suggest an accelerated aging process. Changes in the brain's structure appear to indicate a disruption to the typical aging process in men with PD. Grasping the diverse neurological features in PD patients of different sexes is imperative for devising targeted interventions and treatments that meet their particular needs. By recognizing the accelerated aging-related changes in the brain structure, we can potentially discover novel therapeutic targets (such as neuroprotective strategies, inflammation and oxidative stress, dopaminergic system modulation, and lifestyle and environmental interventions) to decrease the impact of PD on brain health and its progression in this population.

## 5. Conclusion:

The objective of this research was to investigate the structural brain-age variations between males and females afflicted with PD. In order to achieve this objective, we developed a framework for determining brain age using a sample size of 949 individuals who were deemed cognitively healthy. We subsequently evaluated the framework on a group of 373 PD patients who were part of the PPMI dataset at baseline. The results of our simulation analysis revealed that male patients diagnosed with PD experience a hastened process of cerebral aging in contrast to clinically matched female brains. The clinical relevance of Brain-PAD in PD has been demonstrated by its significant correlation with many different clinical variables including UPDRS-III score. Interestingly, a lower MoCA score was significantly associated with Brain-PAD only in male PD while it was not significant in female PD patients. These results may suggest that the PD-related accelerated aging atrophy phenotype is expressed with a significantly higher magnitude in male patients compared to female patients. The expression of this pattern is related to accelerated cognitive decline, particularly in the visuospatial domain, and RBD in male patients.

**Acknowledgments**

Data used in the preparation of this article were obtained from the Open Access Series of Imaging Studies (OASIS), the IXI, and Parkinson's Progression Markers Initiative (PPMI) databases. We wish to thank all investigators and participants of these projects who collected these valuable datasets and made them freely accessible.

The OASIS project is funded by grants P50 AG05681, P01 AG03991, P01 AG026276, R01 AG021910, P20 MH071616, U24 RR021382. Principal Investigators: D. Marcus, R, Buckner, J, Csernansky J. Morris. See http://www.oasis-brains.org/ for more details.

The IXI data used in the preparation of this manuscript were supported by the U.K. Engineering and Physical Sciences Research Council (EPSRC) GR/S21533/02. The IXI Dataset is a collection of nearly 600 MR images from normal, healthy subjects. The MR image acquisition protocol for each subject includes T1, T2 and PD-weighted images, MRA images, and diffusion-weighted images (15 directions). The data was collected at three different hospitals in London using 1.5T and 3T scanners. See http://www.brain-development.org/ for more details.

The PPMI database used in this study is funded by the Michael J. Fox Foundation for Parkinson's Research and funding partners, including AbbVie, Avid Radiopharmaceuticals, Biogen, Bristol-Myers Squibb, Covance, GE Healthcare, Genentech, GlaxoSmithKline (GSK), Eli Lilly and Company, Lundbeck, Merck, Meso Scale Discovery (MSD), Pfizer, Piramal Imaging, Roche, Sanofi Genzyme, Servier, Teva, and UCB. See www.ppmi-info.org/fundingpartners for more details.

**Author Contributions**

Conceptualization, I.B. and J.H.K.; methodology—software—validation—visualization, I.B.; writing—original draft preparation, I.B., and S.B.; supervision, J.H.K. All authors have read and approved and contributed to the final written manuscript.

**Funding**

The authors' research program and/or salary is supported by Natural Science and Engineering Research Council of Canada and Parkinson Canada.

**Data Availability**

Data used in the preparation of this article were obtained from three open-access datasets: the Open Access Series of Imaging Studies (OASIS; http://www.oasis-brains.org), the IXI

(http://www.brain-development.org/), and the Parkinson's Progression Markers Initiative (PPMI; www.ppmi-info.org/data). The data used in this study were downloaded on September 10, 2022.

**Code Availability**

We implemented the brain age estimation and bias adjustment approach using our previously validated framework, which can be accessed from the public repository at https://github.com/medicslab/Bias_Correction.

**Conflicts of Interest**

All authors declare that there is no conflict of interest on their part.